\documentclass[11pt,a4paper]{article}
\usepackage[utf8]{inputenc}
\usepackage{amsmath}
\usepackage{amsfonts}
\usepackage{amssymb}
\usepackage{graphicx}
\usepackage{setspace}
\usepackage{color}
\usepackage{enumitem}

\numberwithin{equation}{section}

\begin{document}

\begin{center}

{\large {\textbf{Information encoded in gene-frequency trajectories}}}

\bigskip

{K. Mavreas and D. Waxman}

{Centre for Computational Systems Biology, ISTBI,\\ 
Fudan University, 220 Handan Road, Shanghai 200433,
PRC\bigskip}

\end{center}

\begin{abstract}
In this work we present a systematic mathematical approximation scheme that exposes the 
way that information, about the evolutionary forces of selection and random genetic drift, 
is encoded in gene-frequency trajectories. 

We determine approximate, time-dependent, gene-frequency trajectory statistics,
assuming additive selection.
We use the probability of fixation to test and illustrate the approximation scheme introduced. 
For the case where the strength of selection and the effective population size 
have constant values, we show how a standard result for the probability of
fixation, under the diffusion approximation, systematically emerges, when increasing 
numbers of approximate trajectory statistics are taken into account. 
We then provide examples of how 
time-dependent parameters influence gene-frequency statistics.
\end{abstract}

\section{Introduction}

A \textit{gene-frequency trajectory}, namely the set of values taken by an
allele's relative frequency over a period of time, encodes information about
the underlying processes that give rise to the trajectory. These processes are
a combination of deterministic and stochastic evolutionary forces.
In the present work, we present a systematic mathematical approximation scheme
that exposes this information.

This work focusses on basic statistics associated with a set of gene-frequency 
trajectories. Such an analysis gives us
the means to understand and quantify how different evolutionary forces influence 
statistics of trajectories and how they feed into quantities of direct interest.

\subsection{Scope of the study}

The primary focus of this work is on the approximation of time-dependent \textit{
trajectory statistics}, which may have various applications. However, we shall
repeatedly apply and test the results we obtain on the \textit{probability of
fixation}, i.e., the probability that an allele \textit{ultimately} achieves a relative
frequency of unity. 

The probability of fixation, which is a quantity of considerable interest in its own right
(see \cite{Kimura} and \cite{EwensBook}), is, for a given initial frequency,
a single number, and constitutes a convenient 
testing ground/target for our approach, compared with trajectory statistics 
such as the mean frequency, which is defined over a range of times, and hence is a function of time, 
and not a single number. 

We consider a single locus in a randomly mating diploid population. While the
evolutionary forces at play in such a population can include mutation, natural selection
and random genetic drift, we shall neglect mutation, assuming that for the timescales/population
sizes considered, mutation occurs with negligible probability.

Natural selection, while often treated as a deterministic force, can have
deterministic and stochastic aspects (see, for example, \cite{Takahata} and
\cite{Takahata_Kimura}). Here, we consider purely deterministic (i.e., predictable)
selection, first with a constant strength, later with a time dependent strength.
Incorporating selection with a stochastic component is possible, within the results
we present, but would require carrying out an average.

Random genetic drift, which we shall sometimes refer to as just `drift' or
`genetic drift', occurs in a population of finite size, and is
stochastic in character \cite{EwensBook}. We will first consider a
constant effective population size, corresponding to genetic drift with a
fixed `strength', and later will allow the effective population size to 
be time dependent, corresponding to drift with a varying strength.

The calculations we present lie in a regime of near neutrality. Thus with $s$
a typical selection coefficient associated with a mutant at the locus of interest, and
$N_{e}$ the effective population size, the regime we consider is
\begin{equation}
N_{e}|s| \lesssim 1. \label{Ne s <1}
\end{equation}
In such a regime we show how it is possible to develop a theoretical
methodology that exposes information about evolutionary forces, that is 
encoded in gene-frequency trajectories.

We note that while the method we present corresponds to a restriction on the
\textit{magnitude} of the selection coefficient of a mutant (Eq.
(\ref{Ne s <1})), it can flexibly deal with selection coefficient of \textit{both signs}. 
This flexibility allows the establishment of results for both
beneficial mutations, which are directly relevant to evolutionary
adaptation, and deleterious mutations, which, for example, play an
important role in the survival of asexual populations (\cite{Muller},
\cite{Felsenstein}). 

As already stated, we primarily test and apply the methods we present on the
probability of fixation. In particular, we show how Kimura's result, for the probability of 
fixation when the strength of selection and the effective population size take constant values, 
emerges as the number of approximate trajectory statistics is increased. We proceed to show how 
to extend the analysis to time dependent parameters.

Because we work in a nearly neutral regime (Eq. (\ref{Ne s <1})), the examples
we give on the fixation probability are complementary to previous results for
this quantity, i.e., for quite strongly beneficial mutations ($4N_{e}s\gg1$)
when the population size is constant (see \cite{Haldane1927}
and, for example, \cite{Mavreas}) or when it changes with time, either monotonically 
\cite{KimuraOhta_N}, or more generally \cite{Otto}), or where selection and population size
change over a finite time \cite{Waxman}), or when selection is very weak 
($N_{e}|s|\ll1$, see e.g., \cite{Lambert}). However, this work is more general than just
an analysis of the fixation probability, since it focuses on \textit{trajectory
statistics}, and in the Discussion some space is devoted to the ways the
methods presented in this work can be extended to a wider class of problems. 

\section{Background}

Consider a randomly mating diploid dioecious sexual population with equal sex ratio.
Generations are discrete, and labelled $0$, $1$, $2$, $\ldots$ . The fitness of each individual is 
determined by a single locus that has two alleles, one of which is a mutant or focal allele, $A$,
and the other a non-focal allele, $B$.

Throughout this work we assume that the phenomena we
consider occur under conditions where new mutations are sufficiently
improbable that they can be neglected. 

We take fitness to be \textit{additive} in nature. The implementation and
parameterisation of additive selection is achieved by writing the relative
fitnesses of the $AA$, $AB$ and $BB$ genotypes as $1+2s$, $1+s$, and $1$,
respectively, where $s$ is a selection coefficient associated with the number
of $A$ alleles in a genotype. While the value of $s$ is restricted in
magnitude, according to Eq. (\ref{Ne s <1}), the sign of $s$ is
unrestricted. Thus $s$ can be positive, negative, or zero, corresponding to mutations
that are beneficial, deleterious, or neutral, respectively. 

Apart from the selection coefficient, another parameter, that plays a key role
in the dynamics, is the effective population size, which we write as $N_{e}$.
This characterises the strength of random frequency fluctuations associated
with random genetic drift. In the simplest case of an ideal population, namely
one described by a Wright-Fisher model (\cite{Fisher}, \cite{Wright}), the
effective population size, $N_{e}$, coincides with the actual (or census)
population size, $N$. However, when there is greater variability in offspring
numbers than that of a Poisson distribution, the effective population size
will be smaller than the census size \cite{EwensBook}, i.e., $N_{e}<N$. Other
deviations from the pure Wright-Fisher model also lead to $N_{e}$ deviating
from $N$ \cite{EwensBook}. The effective population size explicitly appears in
the diffusion equation associated with the diffusion approximation \cite{Rice}
and can be incorporated into simulations of the Wright-Fisher model \cite{ZhaoGossmannWaxman}.

\subsection{Fixation probability}

In the biallelic population described above, which is characterised by the
parameters $s$ and $N_{e}$, a consequence of the neglect of mutation is that 
fixation and loss of the different alleles are the
only possibilities at long times. The probability that the $A$ allele
ultimately achieves fixation, termed the \textit{fixation probability}, was
obtained by Kimura under the diffusion approximation \cite{Kimura}. In terms
of a composite parameter $R$ defined by
\begin{equation}
R=4N_{e}s\label{R def}
\end{equation}
which is a scaled measure of the strength of selection, Kimura found that when
the initial frequency of the $A$ allele is $y$, the probability of fixation is
approximately
\begin{equation}
P_{fix}(y)=\frac{1-e^{-Ry}}{1-e^{-R}}\label{Kimura}
\end{equation}
\cite{Kimura}. Note that this result has the properties that it vanishes at an initial frequency of 
zero ($\lim_{y\rightarrow0} P_{fix}(y)=0$) and it takes the value of unity at an initial 
frequency of unity ($\lim_{y\rightarrow1}P_{fix}(y)=1$).

Kimura's result, in Eq. (\ref{Kimura}), is relatively simple to state, but it
takes some mathematical machinery to derive, requiring, for example, solution
of a backward or forward diffusion equation (see, for example, \cite{Kimura} and
\cite{McKane_Waxman}, respectively). In the present work we also use a
diffusion approximation, but show how results, such as Eq. (\ref{Kimura}), can
simply and systematically \textit{emerge} from the inclusion of some basic
statistics of gene-frequency trajectories.

Indeed, the resulting understanding, that basic properties of fixation and
loss can be derived from essentially elementary statistics of gene frequency
trajectories, allows principled generalisations of results, such as Eq.
(\ref{Kimura}), to more realistic and complex scenarios involving selection
and population sizes that are time-dependent, and we will give examples of this. 
Furthermore, the analysis that we present explicitly illustrates the way that the important
information is encoded in statistics of gene frequency trajectories.

\subsection{What determines gene frequency trajectories?}

We now give some background, that we shall shortly use, on what theoretically
determines gene-frequency trajectories, and hence which underlies, for
example, Kimura's result for the probability of fixation (Eq. (\ref{Kimura})).

To proceed, let $X(t)$ denote the relative frequency (\textit{frequency} for
short) of the $A$ allele at time $t$, with $1-X(t)$ the corresponding
frequency of the $B$ allele. A general feature of the frequency, since it is a
proportion, is that for all $t$ it lies in the range $0$ to $1$, which
includes the end points of the range, namely $0$ and $1$.

We take time to run from an initial value of $0$, and the frequency at this
time, termed the \textit{initial frequency}, is denoted by $y$, i.e.,
\begin{equation}
X(0)=y. \label{initial frequency}
\end{equation}
A particular gene (or allele) frequency trajectory is specified by the form of 
$X(t)$ for a range of times that (in the present work) starts at $t=0$.

We can think of the dynamics of the frequency as being driven by evolutionary
forces, which generally cause changes in the frequency. In the present case,
we have assumed a one locus problem where 
there are only two
evolutionary forces acting, namely selection and random genetic drift. We
shall assume that these two forces are weak, in the absolute sense that, 
from one generation to the next, they cause only small changes in the frequency. We
can then, reasonably, work under the \textit{diffusion approximation}, where time and frequency
are treated as continuous variables. A frequency trajectory is then
approximated as a \textit{continuous} function of \textit{continuous} time.

Over the very small time interval, from $t$ to $t+dt$, the change in the
frequency, $dX(t)=X(t+dt)-X(t)$, derives a contribution from the
systematic forces that are acting (i.e., forces that exclude random
genetic drift). When the frequency is $x$ this change in frequency is written
as $F(x)dt$, with $F(x)$ the systematic force. In the problem at hand, this
force is derived purely from selection. Under additivity, as defined
above, we have, to leading order in $s$, 
\begin{equation}
F(x)=sx(1-x). \label{F(x)}
\end{equation}

When the frequency at time $t$ is $x$, the corresponding contribution to
$dX(t)$ from genetic drift is
\begin{equation}
\sqrt{V(x)}dW(t). \label{drift force}
\end{equation}
The first factor in this expression involves the function
$V(x)$, which is a measure of the variance of allele frequency caused by
drift, and sometimes called the \textit{infinitesimal variance}. The form of $V(x)$
originates in the Wright-Fisher model \cite{EwensBook}, and is given by
\begin{equation}
V(x)=\frac{x(1-x)}{2N_{e}}.
\end{equation}
The other factor in Eq. (\ref{drift force}) is the quantity
$dW(t)=W(t+dt)-W(t)$, which represents the random `noise' associated with
genetic drift\footnote{The quantity $W(t)$ is a \textit{Wiener process} or
\textit{Brownian motion}. It is a random function of the time
with mean zero. For the full set of properties of $W(t)$ see e.g.,
\cite{Tuckwell}).}. Combining the contributions from selection and drift, we
obtain the following differential equation for the change in $X(t)$ from time $t$ to
time $t+dt$:
\begin{equation}
dX(t)=F(X(t))dt+\sqrt{V(X(t))}dW(t). \label{X SDE}
\end{equation}
Equation (\ref{X SDE}) contains randomness\footnote{Formally, Eq. (\ref{X SDE}) is
an \textit{Ito stochastic differential equation} \cite{Ito} and has the key
property that $X(t)$ and $dW(t)$ are statistically independent, as we shall use
later.}
and is one way of representing the diffusion approximation of random genetic
drift. Another way of representing this approximation is in terms of the
equation obeyed by the distribution (probability density) of $X(t)$. It can be
shown that Eq. (\ref{X SDE}) directly leads to the distribution of $X(t)$
obeying a diffusion equation (see, e.g., \cite{Tuckwell}).

Equation (\ref{X SDE}) determines allele frequencies over a range of times,
and so determines \textit{gene frequency trajectories}. Since $F(x)$ and
$V(x)$ have been specified, we can obtain an approximate realisation of a
trajectory by numerically solving Eq. (\ref{X SDE}) over a given time
interval, when starting from an initial frequency of $y$ at time $0$, and
using a particular realisation of the noise from random genetic
drift\footnote{A realisation of the noise corresponds to the specification of
the random function, $W(t)$, over a \textit{range} of times, starting from
time $0$.}. If we solve Eq. (\ref{X SDE}) again, with the same initial
frequency, but with a different realisation of the noise, then we obtain
a different realisation of a frequency trajectory.

\subsection{Statistics of trajectories}

Statistics of trajectories, such as the expected (or mean) value of the
frequency at a given time, $t$, are obtained by averaging $X(t)$ over many
trajectories, and we leave it implicit that every trajectory starts at 
frequency $y$ (see Eq. (\ref{initial frequency})). Generally, we will
indicate such mean values by an overbar, for example, the mean values of
$X(t)$ and $[X(t)]^{2}$ are written as $\bar{X}(t) \equiv\overline{X^{1}}(t)$
and $\overline{X^{2}}(t)$, respectively. At $t=0$ these mean values reduce to $y$ and
$y^{2}$, respectively, because the initial frequency has the definite value $y$.

To illustrate just some of the information that is contained within statistics of
frequency trajectories, let us consider the fixation probability, which, for
an initial frequency of $y$, we write as $P_{fix}(y)$. For any positive
constant, $c$, we can write the fixation probability as a long time limit
\begin{equation}
P_{fix}(y)=\lim_{t\rightarrow\infty}\overline{X^{c}}(t) \label{lim}
\end{equation}
which follows because at long times, the only outcomes are loss or fixation\footnote{Equation (\ref{lim}) follows
since as $t\rightarrow\infty$, the frequency $X(t)$ only achieves one of two
values, namely $0$ (loss) and $1$ (fixation), and it does so with the
probabilities $1-P_{fix}(y)$ and $P_{fix}(y)$, respectively. Consequently
$\lim_{t\rightarrow\infty}\overline{X^{c}}(t)=0^{c}\times[1-P_{fix}
(y)]+1^{c}\times P_{fix}(y)$ and any $c>0$ leads to Eq. (\ref{lim}).}. Thus
knowledge of the mean value of any positive power of $X(t)$, for all $t$, is
fully sufficient to determine the fixation probability.

\section{Approximate trajectory statistics - with time-independent parameters}

In principle, the initial frequency, $y$, and
the differential equation that governs the behaviour of the frequency, $X(t)$ 
(Eqs. (\ref{initial frequency}) and (\ref{X SDE}), respectively), contain 
all available information on frequency trajectories. We cannot, however, directly get access to this
information because Eq. (\ref{X SDE}) cannot, generally, be analytically solved
for $X(t)$ \cite{Kloeden}. We shall proceed by carrying out an approximate
analytical analysis, where we determine approximate time-dependent trajectory statistics,
and in this way gain access to information encoded within trajectories.

As we shall shortly show, a given calculation simultaneously determines a
\textit{set} of approximate time-dependent trajectory statistics. In particular, if a
calculation yields an approximation to $\overline{X^{1}}(t)$, $\overline
{X^{2}}(t)$, $\overline{X^{3}}(t)$, \ldots, $\overline{X^{n}}(t)$, then we say
 `we have an $n$'th order approximation to the problem.' Thus if we
approximately determine just $\overline{X^{1}}(t)$ then we have a
\textit{first order approximation}, while if we approximately determine both
$\overline{X^{1}}(t)$ and $\overline{X^{2}}(t)$ then we have a \textit{second
order approximation}.

By virtue of Eq. (\ref{lim}), an $n$'th order approximation, for any non-zero
$n$, rather directly contains information about the fixation probability. We
shall illustrate the quality and content of the $n$'th order approximation by
comparing results for the fixation probability, for some different values of
$n$. We begin with the first order approximation, i.e., an approximation of
just $\overline{X^{1}}(t)\equiv\bar{X}(t)$.

\subsection{\label{first order section} First order approximation}

Indicating expectations (or mean values) by an overbar, the expected
value of Eq. (\ref{X SDE}) is $d\bar{X}(t)=\overline{F(X(t))}dt+\overline
{\sqrt{V(X(t))}dW}(t)$, and statistical independence of $X(t)$ and $dW(t)$
leads to the second term, on the right hand side of this averaged equation, vanishing\footnote{Omitting
time arguments, we have that $\overline{\sqrt{V(X)}dW}$ equals $\overline{\sqrt{V(X)}}\times
d\overline{W}$ (by statistical independence), which then vanishes because
$d\overline{W}=0$.}. We thus obtain $d\bar{X}(t)=\overline{F(X(t))}dt$ or
\begin{equation}
\frac{d\bar{X}(t)}{dt}=s\left[  \bar{X}(t)-\overline{X^{2}}(t)\right]
.\label{1}
\end{equation}
Equation (\ref{1}) follows from Eq. (\ref{X SDE}) with no approximations, and
while we can say the right hand side of Eq. (\ref{1}) has its \textit{origin} in 
selection, this identification is not completely straightforward\footnote{We
can say the right hand side of Eq. (\ref{1}) has its \textit{origin} in selection, in the
sense that it is the expected value of the selective force $sX(t)\left[
1-X(t)\right]$. However, the right hand side of Eq. (\ref{1}) contains the
expected values $\bar{X}(t)$ and $\overline{X^{2}}(t)$, which are the outcome
of \textit{both} of the evolutionary forces acting within Eq. (\ref{X SDE}). As a
consequence, $\bar{X}(t)$ and $\overline{X^{2}}(t)$ depend on both $s$ and
$N_{e}$ (see later results, e.g., Eq. (\ref{xbar x2bar n=2})), and hence
contain effects of both selection and random genetic drift.}.

We note that purely from the viewpoint of Eq. (\ref{1}), we cannot determine
$\bar{X}(t)$ because the function $\overline{X^{2}}(t)$ is present but
unknown, and Eq. (\ref{1}) gives no information about $\overline{X^{2}}(t)$.
We shall thus pursue approximations.

Two simple first order approximations of Eq. (\ref{1}), which both allow
explicit determination of $\bar{X}(t)$, suggest themselves. These are: (i)
omit $\overline{X^{2}}(t)$, or (ii) replace $\overline{X^{2}}(t)$ by $[\bar
{X}(t)]^{2}$. However, both approximations lead to forms of $\bar{X}(t)$ with
large $t$ behaviours that are unsatisfactory. The large $t$ limit of
$\bar{X}(t)$ of approximation (i) either diverges or vanishes, depending on
the sign of $s$, while that of approximation (ii) either vanishes or is unity,
again depending on the sign of $s$. Neither approximation, when used in Eq.
(\ref{lim}) with $c=1$, leads to a meaningful result for the fixation
probability, which cannot diverge, and generally has a value that lies between
$0$ and $1$.

A preferable first order approximation of Eq. (\ref{1}) is based on noting
that when the time, $t$, gets large, $\bar{X}(t)$ and $\overline{X^{2}}(t)$
both approach the \textit{same} limit, which is an exact property that follows from
Eq. (\ref{lim}). It suggests that $\bar{X}(t)-\overline{X^{2}}(t)$ is not
large for an appreciable amount of time, and motivates the approximation of
simply \textit{omitting} $s[\bar{X}(t)-\overline{X^{2}}(t)]$ in Eq. (\ref{1}),
i.e., omitting the entire right hand side of this equation. This leads to
\begin{equation}
\frac{d\bar{X}(t)}{dt}\simeq0\quad\text{first order approximation.}
\label{first}
\end{equation}
The solution to Eq. (\ref{first}), subject to $X(0)=y$, is simply
\begin{equation}
\bar{X}(t)\simeq y\quad\text{first order approximation.}
\label{first - solution}
\end{equation}

\subsubsection{Fixation probability}

The first order approximation of $X(t)$ in Eq. (\ref{first - solution}) can be
used to approximate the fixation probability. Setting $c=1$ in Eq.
(\ref{lim}), and using Eq. (\ref{first - solution}) leads to the neutral
fixation probability, namely $P_{fix}(y)=\lim_{t\rightarrow\infty}\bar
{X}(t)\simeq y$, that follows from the $s\rightarrow 0$ limit of Eq. (\ref{Kimura}).
For reasons that will shortly become clear, we write this result for the
fixation probability in the form
\begin{equation}
P_{fix}(y)\simeq\frac{\tfrac{(Ry)}{1!}}{\tfrac{R}{1!}}\quad\text{first
order approximation.} \label{Pfix first}
\end{equation}

For small $R$ ($|R|\ll1$), Eq. (\ref{first - solution}) (or Eq.
(\ref{Pfix first})) is a valid approximation of Eq. (\ref{Kimura}).
It also exhibits appropriate $y$ behaviour: the approximation for
$P_{fix}(y)$ takes the exact value $0$ at $y=0$, increases with $y$, and
achieves the exact value $1$ when $y=1$.

Let us proceed to more sophisticated expressions, by considering a second
order approximation.

\subsection{Second order approximation}

From Eq. (\ref{X SDE}) we can determine the following equation for
$\overline{X^{2}}(t)$:
\begin{equation}
\frac{d\overline{X^{2}}(t)}{dt}=2s\left[ \overline{X^{2}}(t)-\overline{
X^{3}}(t)\right] +\frac{1}{2N_{e}}\left[ \bar{X}(t)-\overline{X^{2}}(t)
\right]  \label{2}
\end{equation}
(see Appendix 1 for details), where the first term on the right hand side
originates in selection, while the second term is an average of the
infinitesimal variance, $V(X(t))$, and hence originates in random genetic drift.

From the viewpoint of Eqs. (\ref{1}) and (\ref{2}), we cannot simultaneously
solve these equations for $\bar{X}(t)$ and $\overline{X^{2}}(t)$ because the
function $\overline{X^{3}}(t)$ is present but unknown, and Eqs. (\ref{1})
and (\ref{2}) give no information about this function. We can, again, make an
approximation that provides a feasible way forward, and allows approximate
determination of $\bar{X}(t)$ and $\overline{X^{2}}(t)$.

We proceed by keeping Eq. (\ref{1}) fully intact, but approximate Eq.
(\ref{2}), using similar reasoning to that used when we made the first order
approximation for $\bar{X}(t)$. In particular, in Eq. (\ref{2}), we omit the
selection-originating term %\newline 
$2s\left[  \overline{X^{2}}(t)-\overline{X^{3}
}(t)\right]  $, assuming it to be small. As will become evident, this
approximation applies when $R$ (Eq. (\ref{R def})) is suitably small. From Eq. (\ref{1}) and the
approximated Eq. (\ref{2}), we thus arrive at a pair of coupled differential
equations for $\bar{X}(t)$ and $\overline{X^{2}}(t)$ given by
\begin{equation}
\left.
\begin{array}
[c]{l}
\dfrac{d\bar{X}(t)}{dt}=s\left[  \bar{X}(t)-\overline{X^{2}}(t)\right]  \\
\\
\dfrac{d\overline{X^{2}}(t)}{dt}\simeq\dfrac{1}{2N_{e}}\left[  \bar
{X}(t)-\overline{X^{2}}(t)\right]
\end{array}
\right\}  \text{second order approximation.}\label{second}
\end{equation}
These equations, combined with $\bar{X}(0)=y$ and $\overline{X^{2}}(0)=y^{2}$,
are sufficient to fully determine $\bar{X}(t)$ and $\overline{X^{2}}(t)$ for
all $t>0$. In particular, when $s$ and $N_{e}$ are independent of time, we
show in Appendix 2 that the solution of the set of equations in Eq.
(\ref{second}) can be written as
\begin{align}
& \left.
\begin{array}
[c]{l}
\bar{X}(t)\simeq\dfrac{Ry-\frac{1}{2}\left(  Ry\right)  ^{2}}{R-\frac{1}
{2}R^{2}}-\dfrac{\frac{1}{2}R^{2}y\left(  1-y\right)  }{R-\frac{1}{2}R^{2}
}e^{-\lambda t}\\
\\
\overline{X^{2}}(t)\simeq\dfrac{Ry-\frac{1}{2}\left(  Ry\right)  ^{2}}
{R-\frac{1}{2}R^{2}}-\dfrac{Ry\left(  1-y\right)  }{R-\frac{1}{2}R^{2}
}e^{-\lambda t}
\end{array}
\right\}  \text{second order approximation}\nonumber\\
& \label{xbar x2bar n=2}
\end{align}
where
\begin{equation}
\lambda=\left(  1-\frac{R}{2}\right)  \frac{1}{2N_{e}}.\label{lambda}
\end{equation}

The approximations for $\bar{X}(t)$ and $\overline{X^{2}}(t)$ in Eq.
(\ref{xbar x2bar n=2}) have time dependence which is governed by $\lambda$.
The approximations have the obvious limitation that $\lambda$ must be
\textit{positive} (to avoid a spurious divergence of the solutions at large $t$, which
occurs if $\lambda$ is negative). This is a clear indication that the
approximation applies under restrictions on the range of values of the $R$
parameter. We consider accuracy of the approximation and the range of $R$ in
a section, below, on numerical accuracy of the fixation probability.

In Figure (1) we compare the form of $\bar{X}(t)$ obtained from the second
order approximation, given in Eq. (\ref{xbar x2bar n=2}), with the
corresponding result for the mean trajectory derived from simulations of the Wright-Fisher
model, using an effective size, $N_{e}$, that differs considerably from the census size, $N$ 
\cite{ZhaoGossmannWaxman}. 

\begin{figure}[pth]
\centering
\includegraphics[width=13.5cm]{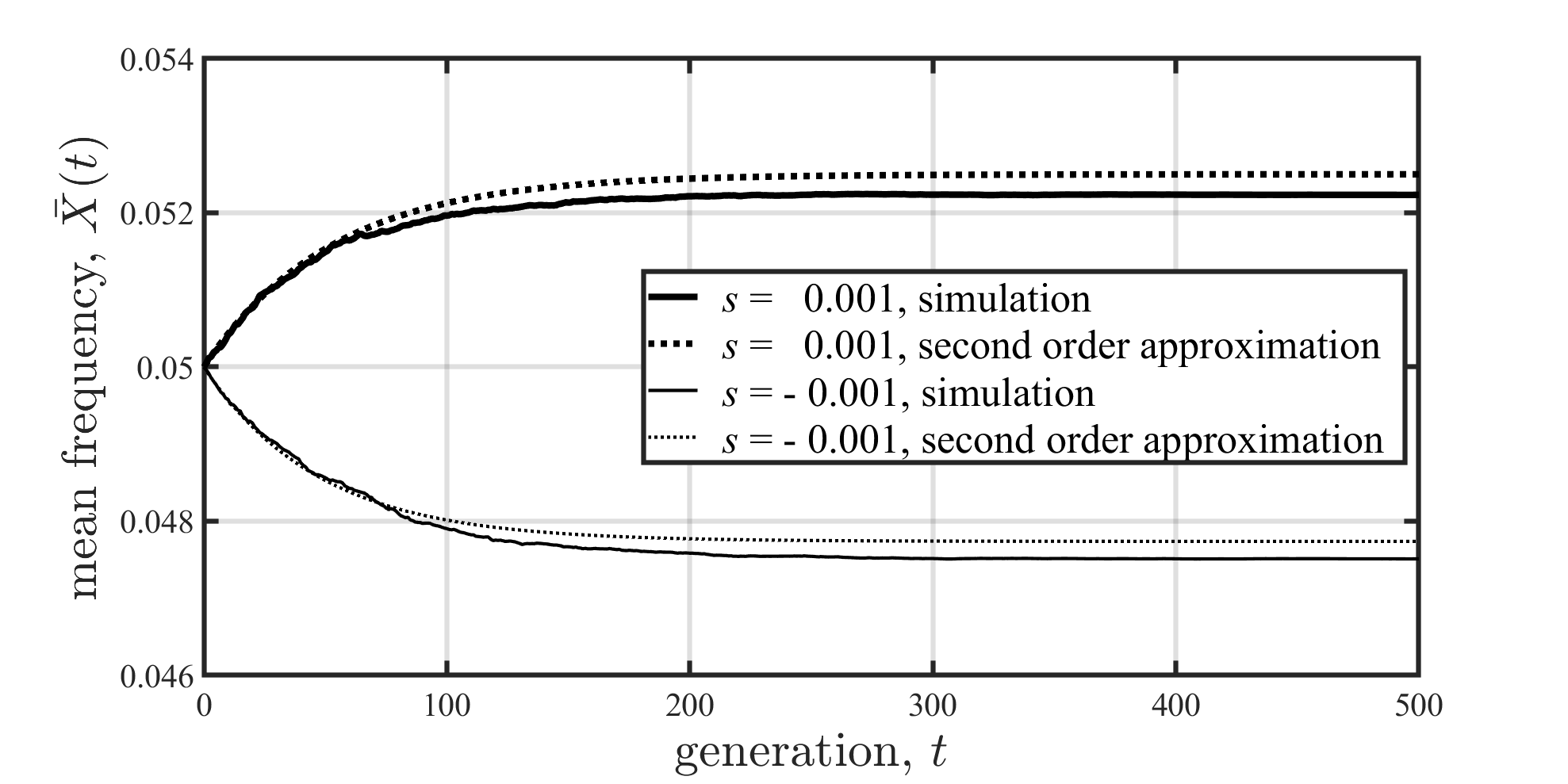} \caption{\textbf{Comparing approximate and simulated mean trajectories.} In this figure we plot
approximate and simulation results for the mean allele frequency, $\bar{X}(t)$,
against the time, $t$.
The approximate results are obtained from the second order
approximation given in Eq. (\ref{xbar x2bar n=2}). The simulation results
are based on a Wright-Fisher model where the effective population size, $N_e$ (that is used in Eq. (\ref{xbar x2bar n=2})) differs from the census size of the population, using the method of \cite{ZhaoGossmannWaxman}. 
The parameter-values adopted are:
census size, $N=100$; effective population size, $N_e=25$; initial frequency, $y=10/200$;
and we give plots for the two selection coefficients $s= \pm 10^{-3}$. The mean of $2 \times 10^6$ 
simulated trajectories were used for each selection coefficient.}
\label{fig:1}
\end{figure}

\newpage

We see from Figure (1) that for the parameter values adopted, there is
reasonable agreement between the different approaches to $\bar{X}(t)$. In
particular, just the second order approximation can capture meaningful time-dependent
features of trajectories.

\subsubsection{Fixation probability}

When $R<2$, the quantity $\lambda$ of Eq. (\ref{lambda}) is positive, and the
approximate forms for $\bar{X}(t)$ and $\overline{X^{2}}(t)$, given in Eq.
(\ref{xbar x2bar n=2}), both converge to the same long time limiting value,
consistent with the long-time limit in Eq. (\ref{lim}) being independent of
the exponent, $c$. This long time limiting value is the second order
approximation of the fixation probability, which is thus given by
\begin{equation}
P_{fix}(y)\simeq\frac{\frac{Ry}{1!}-\frac{(Ry)^{2}}{2!}}{\frac{R}{1!}
-\frac{R^{2}}{2!}}\qquad\text{second order approximation.}
\label{Pfix second}
\end{equation}
We note that, irrespective of the value of $R$, the approximation in Eq.
(\ref{Pfix second}) has the intrinsic feature of taking the exact values $0$
and $1$ when $y$ approaches $0$ and $1$, respectively.

\subsection{Higher order approximations - applied to the fixation probability}

We note that the first and second order approximations of the fixation
probability, given in Eqs. (\ref{Pfix first}) and (\ref{Pfix second}),
respectively, can be seen to follow from the fixation probability of Kimura
(Eq. (\ref{Kimura})), when both numerator and denominator in Kimura's result
are \textit{separately} expanded, to first or second order in $R$, respectively.

We shall use the notation $[\ldots]_{n}$ to denote expansion of the bracketed
quantity to $n$'th order in $R$. For example $[1-e^{-kR}]_{3}$ contains all
terms in $1-e^{-kR}$ up to and including $O(R^{3})$ and is given by
$[1-e^{-kR}]_{3}=\frac{kR}{1!}-\frac{(kR)^{2}}{2!}+\frac{(kR)^{3}}{3!}$. We
can then write Eq. (\ref{Pfix first}) as $P_{fix}(y)\simeq\frac{\left[
1-e^{-Ry}\right]  _{1}}{\left[  1-e^{-R}\right]  _{1}}$, while Eq.
(\ref{Pfix second}) can be written as $P_{fix}(y)\simeq\frac{\left[
1-e^{-Ry}\right]  _{2}}{\left[  1-e^{-R}\right]  _{2}}$. Under a third order
approximation, we consider the set of coupled differential equations for
$\bar{X}(t)$, $\overline{X^{2}}(t)$ and $\overline{X^{3}}(t)$, but now, in the
differential equation for $\overline{X^{3}}(t)$, we omit the term that
originates from selection, namely $3s\left[  \bar{X^{3}}(t)-\overline{X^{4}
}(t)\right]  $, with the resulting equations given in Appendix 3, in Eq.
(\ref{3rd order ode orig params}). This leads to the approximate result
\begin{equation}
\begin{array}
[c]{rclc}
P_{fix}(y) & \simeq & \dfrac{\frac{Ry}{1!}-\frac{(Ry)^{2}}{2!}+\frac{(Ry)^{3}
}{3!}}{\frac{R}{1!}-\frac{R^{2}}{2!}+\frac{R^{3}}{3!}} & \\
&  &  & \\
& \equiv & \dfrac{\left[  1-e^{-Ry}\right]  _{3}}{\left[  1-e^{-R}\right]
_{3}} & \qquad\text{third order approximation}
\end{array}
\label{Pfix third}
\end{equation}
as is shown Appendix 3.

It then becomes highly plausible that an $n$'th order approximation, where
$\overline{X^{1}}(t)$, $\overline{X^{2}}(t)$, \ldots, $\overline{X^{n}}(t)$,
are determined by omitting the selection-originating term in the differential
equation for $\overline{X^{n}}(t)$, leads to
\begin{equation}
P_{fix}(y)\simeq\frac{\left[  1-e^{-Ry}\right]  _{n}}{\left[  1-e^{-R}\right]
_{n}}\qquad\text{}n\text{'th order approximation}\label{Pfixn}
\end{equation}
and this is proved in Appendix 3.

\subsection{\label{accuracy} Numerical accuracy of the fixation probability}

It is of interest to have an indication of the largest value of $|R|$ for
which the $n$'th order approximation works to a given accuracy. This is most
simply found, when applied to the fixation probability, which is a single number 
(for a given initial frequency). To this end, we
introduce a quantity we call $R_{n}(\varepsilon)$, such that for $|R|<R_{n}
(\varepsilon)$ the error on the $n$'th order approximation of the fixation
probability, compared with
Kimura's result\footnote{The error is calculated when all parameters are independent of time,
and applies irrespective of the value of the initial frequency, $y$.}, never exceeds $\varepsilon$.  In Table 1 we give numerically determined values of $R_{n}(\varepsilon)$ for
the orders $n=1$, $2$, \ldots, $5$ of the approximation, and for the error
values $\varepsilon=2\%$, $5\%$, and $10\%$.

\newpage

\begin{table}

\centering

% \caption{\textbf{Error on the approximations.} For time-independent values of the parameters $s$ and
% $N_{e}$, we define the parameter $R_{n}(\varepsilon)$ such that for
% $|R|<R_{n}(\varepsilon)$ the $n$'th order approximation of the fixation
% probability (Eq. (\ref{Pfixn})) has an error, compared with Kimura's result
% (Eq. (\ref{Kimura})), that never exceeds $\varepsilon$. This table contains
% numerically determined values of $R_{n}(\varepsilon)$ for different values of $\varepsilon$ and
% $n$.}
\begin{tabular}
[c]{|cc|c|}\hline
{error, $\varepsilon$} & \multicolumn{1}{|c|}{order, $n$} & $R_{n}(\varepsilon
)$\\\hline
\multicolumn{1}{|c|}{} & $1$ & $0.04$\\\cline{2-3}
\multicolumn{1}{|c|}{} & $2$ & $0.33$\\\cline{2-3}
\multicolumn{1}{|c|}{$2\%$} & $3$ & $0.74$\\\cline{2-3}
\multicolumn{1}{|c|}{} & $4$ & $1.14$\\\cline{2-3}
\multicolumn{1}{|c|}{} & $5$ & $1.56$\\\hline
\multicolumn{1}{|c|}{} & $1$ & $0.10$\\\cline{2-3}
\multicolumn{1}{|c|}{} & $2$ & $0.50$\\\cline{2-3}
\multicolumn{1}{|c|}{$5\%$} & $3$ & $0.99$\\\cline{2-3}
\multicolumn{1}{|c|}{} & $4$ & $1.40$\\\cline{2-3}
\multicolumn{1}{|c|}{} & $5$ & $1.85$\\\hline
\multicolumn{1}{|c|}{} & $1$ & $0.19$\\\cline{2-3}
\multicolumn{1}{|c|}{} & $2$ & $0.68$\\\cline{2-3}
\multicolumn{1}{|c|}{$10\%$} & $3$ & $1.24$\\\cline{2-3}
\multicolumn{1}{|c|}{} & $4$ & $1.62$\\\cline{2-3}
\multicolumn{1}{|c|}{} & $5$ & $2.13$\\\hline
\end{tabular}

\caption{\textbf{Error on the approximations.} For time-independent values of the parameters $s$ and
$N_{e}$, we define the parameter $R_{n}(\varepsilon)$ such that for
$|R|<R_{n}(\varepsilon)$ the $n$'th order approximation of the fixation
probability (Eq. (\ref{Pfixn})) has an error, compared with Kimura's result
(Eq. (\ref{Kimura})), that never exceeds $\varepsilon$. This table contains
numerically determined values of $R_{n}(\varepsilon)$ for different values of $\varepsilon$ and
$n$.}

\end{table}

\newpage

One example of the usage of Table 1 is when $|R|$ has a value less than $1$.
Then a third order approximation leads to an error in the fixation probability
that deviates from Kimura's result by less than $5\%$. However, the real use of Table 1 
is in more complex situations, as we consider next.

\section{Approximate trajectory statistics - with time-dependent parameters}

We shall now use the methodology, developed above, to determine results for
the case where parameters depend on time. This seems to be a principled way to
proceed, since, as we have seen, there is a systematic development of results
such as the fixation probability, with the order of approximation.

Let us reconsider the model considered above, but
now with the parameters $s$ and $N_{e}$ varying with time, i.e., with $s=s(t)$ and
$N_{e}=N_{e}(t)$. This leads to the composite parameter $R$ becoming a
function of time, i.e., $R(t)=4N_{e}(t)s(t)$.

We shall proceed under the assumption that from $t=0$ onwards, the value of
$|R(t)|$ remains small. For example if $|R(t)|$ is always below $0.5$ then the
results in Table 1 make it \textit{plausible} that if we use the second order
($n=2$) approximation, there will be an error that is smaller than $5\%$ in
the result obtained for the fixation probability.

Since a first order approximation, is, by Eq. (\ref{first - solution}),
independent of any parameters, we shall consider non-trivial cases of second
and third order approximations.

\subsection{Second order approximation - with time-dependent parameters}

For the second order approximation, the functions $\bar{X}(t)$ and
$\overline{X^{2}}(t)$ continue to approximately satisfy Eq. (\ref{second}) but
now the quantities $s$ and $N_{e}$, that are present in the equations, are
time-dependent. There are various ways to write the solutions for $\bar{X}
(t)$, $\overline{X^{2}}(t)$ and $P_{fix}(y)$. One such way is in terms of a
function $\Phi(t)$ defined by
\begin{equation}
\Phi(t)=1-\exp\left(  -\int_{0}^{t}\left(  1-\frac{R(z)}{2}\right)
\frac{dz}{2N_{e}(z)}\right)  .\label{Phi(t)}
\end{equation}
Then with $\Phi^{\prime}(t)=d\Phi(t)/dt$ we find we can write
\begin{equation}
\bar{X}(t)\simeq\int_{0}^{t}\frac{\left[  1-e^{-R(z)y}\right]  _{2}}{\left[
1-e^{-R(z)}\right]  _{2}}\Phi^{\prime}(z)dz+y[1-\Phi
(t)].\label{xbar t dep parameters}
\end{equation}
(see Appendix 4 for details). The second order approximation for
$\overline{X^{2}}(t)$ follows from Eq. (\ref{xbar t dep parameters}) by
replacing $y$ by $y^{2}$ in the factor multiplying $[1-\Phi(t)]$.

We note that the form of $\bar{X}(t)$ in Eq. (\ref{xbar t dep parameters}) has
an apparent probabilistic interpretation\footnote{To see the probabilistic
interpretation of Eq. (\ref{xbar t dep parameters}), we
introduce a random variable $\tau$ with cumulative probability distribution
$\operatorname*{Prob}(\tau\leq t)=\Phi(t)$ and probability density
$\Phi^{\prime}(t)=d\Phi(t)/dt$. Then the form of $\bar{X}(t)$ in Eq.
(\ref{xbar t dep parameters}) coincides with the average of a function that,
for $\tau\leq t$, 
takes the value $\frac{\left[  1-e^{-R(\tau)y}\right]_{2}}{\left[
1-e^{-R(\tau)}\right]_{2}}$,  and for $\tau
>t$, takes the value $y$.}.

It may be verified that when $s$ and $N_{e}$ are independent of time, Eq.
(\ref{xbar t dep parameters}) reduces to Eq. (\ref{xbar x2bar n=2}).

Since we work under the assumption of relatively small $|R(t)|$ (i.e.,
$|R(t)|\lesssim1$) it follows that as $t\rightarrow\infty$ we have
$\Phi(t)\rightarrow1$, hence from $P_{fix}(y)=\lim_{t\rightarrow\infty}\bar
{X}(t)$ and from Eq. (\ref{xbar t dep parameters}) we obtain the approximate
result
\begin{align}
P_{fix}(y)  &  \simeq\int_{0}^{\infty}\frac{\left[  1-e^{-R(t)y}\right]  _{2}
}{\left[  1-e^{-R(t)}\right]  _{2}}\Phi^{\prime}(t)dt\nonumber\\
& \nonumber\\
&  =y^{2}+y(1-y)\int_{0}^{\infty}e^{-\int_{0}^{t}\left(  1-\tfrac{R(z)}
{2}\right)  \tfrac{dz}{2N_{e}(z)}}\frac{dt}{2N_{e}(t)}.
\label{Pfix2 time dependent}
\end{align}

The first form of the fixation probability in Eq. (\ref{Pfix2 time dependent})
is equivalent to an average of $\frac{\left[  1-e^{-R(t)y}\right]
_{2}}{\left[  1-e^{-R(t)}\right]  _{2}}$, with $\Phi^{\prime}(t)$ playing the
role of a probability density. This tells us, without any additional
calculation, that the approximation of $P_{fix}(y)$ in Eq.
(\ref{Pfix2 time dependent}) lies between the smallest and largest values that
$\frac{\left[  1-e^{-R(t)y}\right]  _{2}}{\left[1-e^{-R(t)}\right]  _{2}}$
takes, from time $t=0$ onwards.

The second form given for the fixation probability in Eq.
(\ref{Pfix2 time dependent}) may be more useful for practical computations.

\subsubsection{Piecewise constant variation}

As a simple illustration of the use of Eq. (\ref{Pfix2 time dependent}),
suppose the effective population size, $N_{e}$, stays constant,
\begin{equation}
N_{e}=N_{0}
\end{equation}
while the selection coefficient changes over time according to
\begin{equation}
s(t)=\left\{
\begin{array}
[c]{lll}
s_{0} &  & \text{for }0 \le t<T\\
&  & \\
s_{1} &  & \text{for } t\geq T.
\end{array}
\right.  \label{s(t) specific}
\end{equation}
In terms of the composite parameters
\begin{eqnarray}
R_{0} &=&4N_{0}s_{0},\quad R_{1}=4N_{0}s_{1},\quad w=\exp \left[ -\left( 1-
\tfrac{R_{0}}{2}\right) \tfrac{T}{2N_{0}}\right]   \notag \\
&&  \notag \\
&&
\end{eqnarray}
we obtain
\begin{equation}
P_{fix}(y)\simeq(1-w)\frac{\left[  1-e^{-R_{0}y}\right]  _{2}}{\left[
1-e^{-R_{0}}\right]  _{2}}+w\frac{\left[  1-e^{-R_{1}y}\right]  _{2}}{\left[
1-e^{-R_{1}}\right]  _{2}}. \label{Pfix time dep s}
\end{equation}
The result in Eq. (\ref{Pfix time dep s}) is a weighted average of approximate
fixation probabilities associated with the selection coefficients of $s_{0}$
and $s_{1}$. The weighting factor, $w$, is determined by the time that the
change in selection coefficient occurs, along with the parameter-values that
apply prior to this change, namely $s_{0}$ and $N_{0}$. A very similar
`weighted average' result also occurs when the effective population size
discontinuously changes, while the selection coefficient stays constant.

\subsection{Third order approximation - with time-dependent parameters}

A third order approximation corresponds to the solution of the three equations
given in Eq. (\ref{3rd order ode orig params}) or Eq. (\ref{3rd order ode}).
However, to extract, e.g., $\bar{X}(t)$ there is a simpler way of proceeding.
We first define two functions $A(t)$ and $B(t)$ via
\begin{equation}
A \equiv A(t)=\bar{X}(t)-\overline{X^{2}}(t),\quad B\equiv B(t)=\overline{
X^{2}}(t)-\overline{X^{3}}(t).
\end{equation}
These are shown in Appendix 5 to obey
\begin{equation}
\left. 
\begin{array}{rcl}
\dfrac{dA}{dt} & \simeq  & -\dfrac{1}{2N_{e}}\left[ \left( 1-\dfrac{R}{2}
\right) A+RB\right]  \\ 
&  &  \\ 
\dfrac{dB}{dt} & \simeq  & -\dfrac{1}{2N_{e}}\left[ -A+\left( 3-R\right) B
\right] 
\end{array}
\right\}   \label{AB system}
\end{equation}
and are subject to $A(0)=y-y^{2}$ and $B(0)=y^{2}-y^{3}$. The third order
problem only requires determination of $A(t)$ and $B(t)$, with statistics of
frequencies following by integration, e.g.,
\begin{equation}
\bar{X}(t)\simeq y+\int_{0}^{t}\frac{R(z)}{4N_{e}(z)}A(z)dz
\label{Xbar t dep n=3}
\end{equation}
(see Appendix 5 for details).

%\subsubsection{Examples}

For $N_{e}$ and $s$ independent of $t$ we can explicitly solve Eq.
(\ref{AB system}). Here, we shall illustrate the working of the above in the
time-dependent case by determining $\bar{X}(t)$ for the specific forms of
$s(t)$ and $N_{e}(t)$ in two different examples.

\subsubsection{Example 1: $s$ constant, $N_{e}$ changing}

We take a constant selection coefficient
\begin{equation}
s = s_{0} \label{s example 1}
\end{equation}
and the time-dependent effective population size
\begin{equation}
N_{e}(t)=N_{0} \times\left\{
\begin{array}
[c]{lll}
1 &  & \text{for } 0 \leq t < T\\
&  & \\
t/T &  & \text{for }T\leq t < 2T\\
&  & \\
2 &  & \text{for }t\geq 2T.
\end{array}
\right.  \label{N example 1}
\end{equation}
In Figure 2 we give the results of numerically solving Eq. (\ref{AB system}) for this
example, and illustrate the form of $\bar{X}(t)$, as derived from Eq.
(\ref{Xbar t dep n=3}).

\begin{figure}[pth]
\centering
\includegraphics[width=13.5cm]{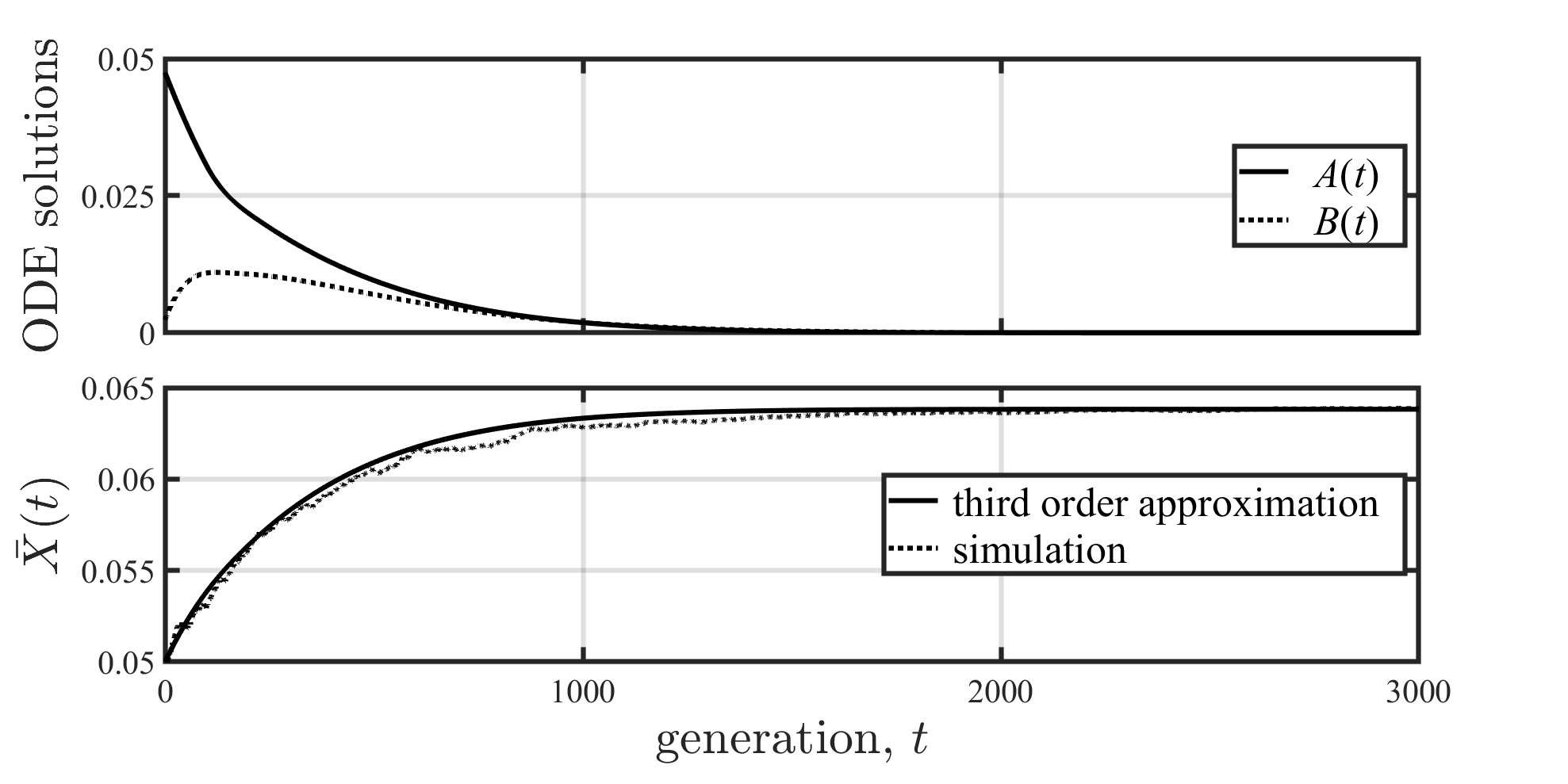}\caption {\textbf{Third order approximation, Example 1.} This figure illustrates results of the third order approximation that follow by first solving the coupled equations given in Eq. (\ref{AB system}) for Example 1, where the selection
coefficient is constant (Eq. (\ref{s example 1})), and the effective population size 
changes over time according to Eq. (\ref{N example 1}). 
The parameter values adopted are $s_{0}=0.001$, $N_{0}=1/100$, $T=100$, and
$y=10/(2N_{0})=1/200$. We plot $A(t)$ (solid line) and $B(t)$ (broken line) against the time, $t$ (top
panel). In the bottom panel we plot the mean frequency,
$\bar{X(t)}$, based on the third order approximation given in Eq.(\ref{Xbar t dep n=3}) (solid line), 
and the mean of $3 \times 10^4$ simulated trajectories (broken line).}
\label{fig:2}
\end{figure}
\newpage

\subsubsection{Example 2: $N_{e}$ constant, $s$ changing}

We take a constant effective population size
\begin{equation}
N_{e}=N_{0}\label{Ne example 2}
\end{equation}
and the time-dependent selection coefficient
\begin{equation}
s(t)=s_{0}\times\left\{
\begin{array}
[c]{lll}
1 &  & \text{for }0\leq t < T\\
&  & \\
t/T &  & \text{for }\leq t < 2T\\
&  & \\
2 &  & \text{for }t \geq 2T.
\end{array}
\right.  \label{s example 2}
\end{equation}
In Figure 3 we give the results of solving Eq. (\ref{AB system}) for this
example, and illustrate the form of $\bar{X}(t)$, as derived from Eq.
(\ref{Xbar t dep n=3}).

\begin{figure}[pth]
\centering
\includegraphics[width=13.5cm]{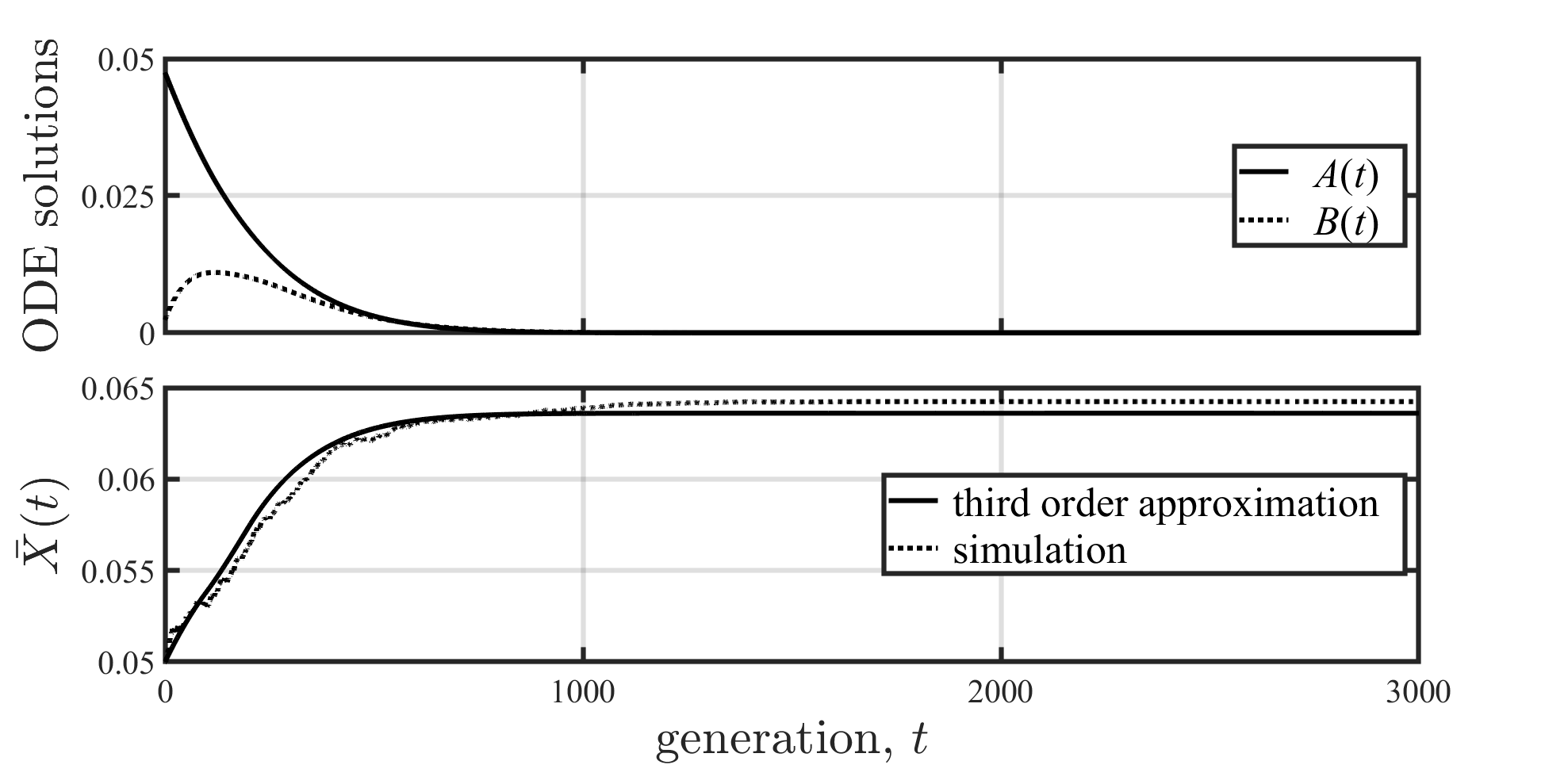} \caption{\textbf{Third order approximation, Example 2.}  
This figure illustrates 
results of the third order approximation that follow by first solving the coupled equations given in Eq. (\ref{AB system}) for Example 2, where the effective population size is constant (Eq. (\ref{Ne example 2})), 
and the selection coefficient changes over time according to Eq. (\ref{s example 2}). 
The parameter values adopted are $s_{0}=0.001$, $N_{0}=1/100$, $T=100$,
and $y=1/(2N_{0})=10/200$. 
We plot $A(t)$ (solid line) and $B(t)$ (broken line) against the time, $t$ (top
panel).
In the bottom panel we plot the mean frequency,
$\bar{X(t)}$, based on the third order approximation given in Eq.(\ref{Xbar t dep n=3}) (solid line), 
and the mean of $3 \times 10^4$ simulated trajectories (broken line).}
\label{fig:3}
\end{figure}
\newpage

In Example 2 the values of $R(t)$ are identical to those in Example 1.
However, the fixation probability generally yields different results when $s$
varies at fixed $N_{e}$, compared with when $N_{e}$ varies at fixed $s$. As a
consequence, despite $R(t)$ taking the same form in both examples, we find a
small but significant difference between the long time values of $\bar{X}(t)$
in the two examples, signalling different fixation probabilities in the two
cases. The effects of genetic drift is different in the two examples (cf.
\cite{Waxman}).

\section{Discussion}
In this work we have presented a systematic mathematical approximation 
scheme that can show approximate parameter-dependencies of statistics of gene-frequency 
trajectories, thus exposing information contained in the trajectories.

The analysis is restricted to a nearly neutral (or weak selection) regime
(see Eq. (\ref{Ne s <1})). It can, however, capture properties
of both negative as well as positive selection coefficients.

We have presented examples for the fixation probability, which is
a long time limit of a trajectory statistic (see Eq. (\ref{lim})).
The reasonable accuracy of the fixation probability, which is related to the mean trajectory, see Figure 1 for results under the second order approximation, suggests that the 
approximations presented have the capability of connecting features of trajectories at early 
and late times.

We note that despite testing and applying the methods presented on the probability of fixation,
it is the case that time dependent frequency trajectory statistics contain
information on more than just this probability. For
example, with $P_{fix}(t,y)$ the probability of fixation \textit{by} time $t$,
given an initial frequency of $y$ at time $0$, it can be shown that for any positive $k$,
\begin{equation}
\overline{X^{k}}(t)\geq P_{fix}(t,y)
\end{equation}
with equality only holding when $k\rightarrow\infty$. Furthermore, from Eq.
(\ref{lim}), we have $\overline{X^{k}}(\infty)=P_{fix}(y)$ and, with $T$
denoting the random time it takes for fixation to be achieved (given that fixation ultimately
occurs), we can write $\operatorname*{Prob}(T\leq
t)=P_{fix}(t,y)/P_{fix}(y)$ and hence
\begin{equation}
\operatorname*{Prob}(T\leq t)\leq\frac{\overline{X^{k}}(t)}{\overline{X^{k}
}(\infty)}.    \label{P(T<t)}
\end{equation}
We can express the mean time to fixation as $\overline{T}=\int_{0}^{\infty}\left[1-\operatorname*{Prob}(T\leq
t)\right]  dt$ and using Eq. (\ref{P(T<t)}) we obtain
\begin{equation}
\overline{T}\geq\int_{0}^{\infty}\left(  1-\frac{\overline{X^{k}}
(t)}{\overline{X^{k}}(\infty)}\right)  dt.\label{ET}
\end{equation}
This result holds for $k=1,2,\ldots$. It thus follows that the particular
\textit{way} that time dependent statistics, such as  $\overline{X^{k}}(t)$,
approach their long time asymptotic values contains information about the
random time to fixation.

We cannot guarantee the inequality in Eq. (\ref{ET}), when we use an
approximate form for $\overline{X^{k}}(t)$, however, we can use it get an \textit{indication} of the
mean time to fixation by determining the right hand side of Eq. (\ref{ET}) e.g., for
$k=2$, using the second order results in Eq. (\ref{xbar x2bar n=2}). We find\\
$\int_{0}^{\infty}\left[  1-\overline{X^{2}}(t)/\overline{X^{2}} (\infty)\right]  dt$ has the approximate value $2N_{e}(1-y)/[(1-R/2)(1-Ry/2)]$. The
neutral diffusion result, for small $y$ is approximately $4N_{e}$ generations,
hence even for $k=2$ we are close to $50\%$ of the full $k= \infty$ result.

The results we have presented in this work are an attempt to extract information 
from the stochastic differential equation $dX(t)=F(X(t))dt+\sqrt{V(X(t))}dW(t)$
that the gene frequency, $X(t)$, obeys. This equation is, generally, not analytically soluble,
despite representing a very simple class of problems, namely those with one locus, two alleles,
additive selection, no mutation, and a finite population size. More complex problems, such as 
those involving non-additive and/or frequency-dependent selection,
mutation/migration, two or more loci, and multiple alleles, are further removed from having analytical 
solution. However, the methodology we have introduced may give some systematic access to such problems.

\newpage

\begin{center}

{\large {\textbf{APPENDICES}}}

\end{center}

%%%%%%%%%%%%%%%%%%%%%%%%%%%%%%%%%%%%%%%%%%%%%%%%%%%%%%%%%%%%
%%% APPENDICES
%%%%%%%%%%%%%%%%%%%%%%%%%%%%%%%%%%%%%%%%%%%%%%%%%%%%%%%%%%%%

\appendix
%\begin{appendixbox}
\section{The equation obeyed by $\overline{X^{n}}(t)$ for $n>1$}

In this appendix we derive the differential equation that $\overline{X^{n}
}(t)$ obeys, where an overbar denotes an expected (or average) value.

Note that we assume that $X(0)$ takes the definite value $y$, thus for all $n$
we have
\begin{equation}
\overline{X^{n}}(0)=y^{n}. \label{ic}
\end{equation}
We begin with the \textit{stochastic differential equation} (SDE) for the
frequency, which is given by
\begin{equation}
dX(t)=F(X(t))dt+\sqrt{V(X(t))}dW(t). \label{Ito}
\end{equation}
This is an Ito SDE \cite{Tuckwell}, which means that $X(t)$ and $dW(t)$ are
statistically independent. Thus the expected value of
$\sqrt{V(X(t))}dW(t)$ equals $\overline{\sqrt{V(X(t))}} \times \overline{dW}(t)$. This vanishes because
$\overline{dW}(t)=0$. Thus Eq. (\ref{Ito}) yields
\begin{equation}
\overline{\sqrt{V(X(t))}dW(t)}=0.
\end{equation}
The expected value of Eq. (\ref{Ito}) then yields $d\bar{X}(t)=\overline
{F(X(t))}dt$ or
\begin{equation}
\frac{d\bar{X}(t)}{dt}=\overline{F(X(t))}. \label{EX1}
\end{equation}

In this work we look at expected value of various powers of $X(t)$. There are
different ways to proceed, but a direct approach derives, from Eq.
(\ref{Ito}), equations involving the expected values of $X^{2}(t)$, $X^{3}
(t)$, \ldots that are analogous to Eq. (\ref{EX1}). The key point is that
the noise increment, $dW(t)$, has an expected value of zero, but behaves
as a random variable with mean $0$ and standard deviation $\sqrt{dt}$. Thus
changes of quantities over a time interval of $dt$, arise 
from terms that are of first order in $dt$, and also from a term that is
second order in $dW(t)$. This is codified in the rules of
Ito calculus \cite{Tuckwell}. With $n=2,3, \ldots$, Ito's rules lead to
a change in $X^{n} $, from $t$ to $t+dt$ (omitting time arguments), of
\begin{align}
dX^{n}  & =nX^{n-1}dX+\frac{1}{2}n(n-1)X^{n-2}\left[  dX\right]
^{2}\nonumber\\
& \nonumber\\
& =nX^{n-1}\left[  F(X)dt+\sqrt{V(X)}dW\right]  +\frac{1}{2}n(n-1)X^{n-2}
V(X)dt.
\end{align}
On averaging this equation, the $dW$ term vanishes, and we arrive at
\begin{equation}
\frac{d\overline{X^{n}}}{dt}=n\overline{X^{n-1}F(X)}+\frac{1}{2}
n(n-1)\overline{X^{n-2}V(X)}. \label{general general}
\end{equation}
When $F(x)=sx(1-x)$ and $V(x)=\frac{1}{2N_{e}}x(1-x)$, Eq.
(\ref{general general}) becomes
\begin{equation}
\frac{d\overline{X^{n}}}{dt}=ns\left(  \overline{X^{n}}-\overline{X^{n+1}
}\right)  +\frac{n(n-1)}{4N_{e}}\left(  \overline{X^{n-1}}-\overline{X^{n}
}\right)  . \label{general}
\end{equation}
For the special case of $n=2$, Eq. (\ref{general}) reduces to
\begin{equation}
\frac{d\overline{X^{2}}}{dt}=2s\left(  \overline{X^{2}}-\overline{X^{3}}
\right)  +\frac{1}{2N_{e}}\left(  \bar{X}-\overline{X^{2}}\right)
\end{equation}
which is Eq. (\ref{2}) of the main text.

%\end{appendixbox}

%\begin{appendixbox}
\section{Solution of $\bar{X}(t)$ and $\overline{X^{2}}(t)$}
In this appendix we determine the form $\bar{X}(t)$ and $\overline{X^{2}}(t)$,
under a second order approximation when the parameters $s$ and $N_{e}$ have
constant values (i.e., are independent of the time).

The equations that $\bar{X}(t)$ and $\overline{X^{2}}(t)$ approximately obey,
under a second order approximation, are given in Eq. (\ref{xbar x2bar n=2}) of
the main text. For convenience we reproduce them here:
\begin{equation}
\left.
\begin{array}
[c]{l}
\dfrac{d\bar{X}(t)}{dt}=s\left[  \bar{X}(t)-\overline{X^{2}}(t)\right]  \\
\\
\dfrac{d\overline{X^{2}}(t)}{dt}\simeq\dfrac{1}{2N_{e}}\left[  \bar
{X}(t)-\overline{X^{2}}(t)\right]
\end{array}
\right\}  \label{second order system}
\end{equation}
Since $X(0)$ takes the definite value $y$, the above equations are subject to
$\bar{X}(0)=y$ and $\overline{X^{2}}(0)=y^{2}$.

There are many ways to solve Eq. (\ref{second order system}), and we adopt the
following approach.

Define
\begin{equation}
D(t)=\bar{X}(t)-\overline{X^{2}}(t)
\end{equation}
then on subtracting the second equation from the first, in Eq.
(\ref{second order system}), we obtain
\begin{align}
\frac{dD(t)}{dt}  & =\left(  s-\dfrac{1}{2N_{e}}\right)  D(t)\nonumber\\
& \nonumber\\
& =-\lambda D(t)\label{D ode}
\end{align}
where $\lambda=\dfrac{1}{2N_{e}}-s$ and using $R=4N_{e}s$ (Eq. (\ref{R def})),
we have
\begin{equation}
\lambda=\left(  1-\frac{R}{2}\right)  \frac{1}{2N_{e}}.
\end{equation}
The solution to Eq. (\ref{D ode}) is $D(t)=D(0)e^{-\lambda t}$ i.e.,
\begin{equation}
D(t)=y(1-y)e^{-\lambda t}.    \label{D}
\end{equation}

The two equations in Eq. (\ref{second order system}) can be written as
$d\bar{X}(t)/dt=sD(t)$ and $d\overline{X^{2}}(t)/dt\simeq D(t)/(2N_{e})$,
respectively. Given the explicit form of $D(t)$ in Eq. (\ref{D}) we can determine
$\bar{X}(t)$ and $\overline{X^{2}}$ by direct integration. The results can be written
as
\begin{equation}
\bar{X}(t)=\frac{Ry-\frac{1}{2}\left(  Ry\right)  ^{2}}{R-\frac{1}{2}R^{2}
}-\frac{\frac{1}{2}R^{2}y\left(  1-y\right)  }{R-\frac{1}{2}R^{2}}e^{-\lambda
t}
\end{equation}
and
\begin{equation}
\overline{X^{2}}(t)=\frac{Ry-\frac{1}{2}\left(  Ry\right)  ^{2}}{R-\frac{1}
{2}R^{2}}-\frac{Ry\left(  1-y\right)  }{R-\frac{1}{2}R^{2}}e^{-\lambda t}.
\end{equation}
Note that these (approximate) forms for $\bar{X}(t)$ and $\overline{X^{2}
}(t)\ $have a number of exact properties:
\begin{enumerate}[label=(\roman*)]
    \item $\bar{X}(0)=y$,
    \item $\overline{X^{2}}(0)=y^{2}$,
    \item $\lim_{y\rightarrow 0}\bar{X}(t)=0$,
    \item $\lim_{y\rightarrow 0}\overline{X^{2}}(t)=0$,
     \item $\lim_{y\rightarrow 1}\bar{X}(t)=1$,
    \item $\lim_{y\rightarrow 1}\overline{X^{2}}(t)=1$,
    \item they have the same long time limiting values (providing $\lambda>0$): \newline
    $\lim_{t\rightarrow\infty}\bar{X}(t)=\lim
_{t\rightarrow\infty}\overline{X^{2}}(t)$.
\end{enumerate}

%\end{appendixbox}

%\begin{appendixbox}
\section{Higher order approximations for the fixation probability}
In this appendix we show how higher order approximations for the fixation
probability can be obtained in the case where the parameters $s$ and $N_{e}$
have constant values.

We begin with a third order approximation.

From Eq. (\ref{general general}) for $n=3$ we have
\begin{equation}
\frac{d\overline{X^{3}}(t)}{dt}=3s\left[  \overline{X^{3}}(t)-\overline{X^{4}
}(t)\right]  +\frac{3} {2N_{e}}\left[  \overline{X^{2}}(t)-\overline{X^{3}
}(t)\right]  . \label{3}
\end{equation}
Proceeding as before, we now omit the assumed small term that originates in
selection in Eq. (\ref{3}), namely $3s\left[  \overline{X^{3}}(t)-\overline
{X^{4}}(t)\right]  $. The resulting approximate equation, along with Eqs.
(\ref{1}) and (\ref{2}), are
\begin{equation}
\left. 
\begin{array}{rcl}
\frac{d\bar{X}(t)}{dt} & = & s\left[ \bar{X}(t)-\overline{X^{2}}(t)\right] 
\\ 
&  &  \\ 
\frac{d\overline{X^{2}}(t)}{dt} & = & \frac{1}{2N_{e}}\left[ \bar{X}(t)-
\overline{X^{2}}(t)\right]  \\ 
&  &  \\ 
&  & +2s\left[ \overline{X^{2}}(t))-\overline{X^{3}}(t)\right]  \\ 
&  &  \\ 
\frac{d\overline{X^{3}}(t)}{dt} & \simeq  & \frac{3}{2N_{e}}\left[ \overline{
X^{2}}(t)-\overline{X^{3}}(t)\right] 
\end{array}
\right\}   \label{3rd order ode orig params}
\end{equation}
and can be written as
\begin{equation}
\left. 
\begin{array}{rcl}
\frac{d\bar{X}(t)}{dt} & = & s\left[ \bar{X}(t)-\overline{X^{2}}(t)\right] 
\\ 
&  &  \\ 
\frac{d\overline{X^{2}}(t)}{dt} & = & \frac{2s}{R}\left[ \bar{X}(t)-
\overline{X^{2}}(t)\right]  \\ 
&  &  \\ 
&  & +2s\left[ \overline{X^{2}}(t))-\overline{X^{3}}(t)\right]  \\ 
&  &  \\ 
\frac{d\overline{X^{3}}(t)}{dt} & \simeq  & \frac{6s}{R}\left[ \overline{
X^{2}}(t)-\overline{X^{3}}(t)\right] 
\end{array}
\right\}   \label{3rd order ode}
\end{equation}
These three equations constitute a closed system that allows determination of
$\bar{X}(t)$, $\overline{X^{2}}(t)$ and $\overline{X^{3}}(t)$.

However, with the parameters $s$ and $N_{e}$ independent of time, we can
determine the fixation probability without explicitly solving for $\bar{X}
(t)$, $\overline{X^{2}}(t)$ and $\overline{X^{3}}(t)$. Rather, we eliminate
$\bar{X}(t)-\overline{X^{2}}(t)$ and $\overline{X^{2}}(t)-\overline{X^{3}}(t)$
from Eq. (\ref{3rd order ode}) to obtain the single equation
\begin{equation}
R\frac{d\bar{X}(t)}{dt}-\frac{R^{2}}{2!}\frac{d\overline{X^{2}}(t)}{dt}
+\frac{R^{3}}{3!}\frac{d\overline{X^{3}}(t)}{dt}\simeq0. \label{rewrite}
\end{equation}
We then integrate Eq. (\ref{rewrite}) over $t$, from $0$ to $\infty$, use Eq.
(\ref{ic}), and identify $\overline{X^{n}}(\infty)$, for $n>0$, with the
fixation probability, $P_{fix}(y)$. We obtain %\newline 
$\left(  R-\frac{R^{2!}}
{2}+\frac{R^{3}}{3!}\right)  P_{fix}(y)-\left(  Ry-\frac{R^{2}y^{2}}{2!}
+\frac{R^{3}y^{3}}{3!}\right)  \simeq0$ which immediately leads to
\begin{equation}
P_{fix}(y)\simeq\frac{Ry-\frac{R^{2}y^{2}}{2!}+\frac{R^{3}y^{3}}{3!}}
{R-\frac{R^{2}}{2!}+\frac{R^{3}}{3!}}. \label{Pfix app appendix}
\end{equation}
This expression can be written as $P_{fix}(y) \simeq\frac{[1-e^{-Ry}]_{3}
}{[1-e^{-R}]_{3}}$, in which: (i) the numerator consists of the leading three
terms of the Taylor series expansion, in $R$, of the numerator of Kimura's
result $P_{fix}(y)=\frac{1-e^{-Ry}}{1-e^{-R}}$, and (ii) the denominator
consists of the leading three terms of the Taylor series expansion, in $R$, of
the denominator of Kimura's result.

We can now show that the $n$'th order approximation of Kimura's fixation
probability is $\frac{[1-e^{-Ry}]_{n}}{[1-e^{-R}]_{n}}$. To obtain this we
begin with Eq. (\ref{general}), and omit the term originating in selection. We
can write this approximate equation, along with the exact forms of Eq.
(\ref{general}) when applied to $\overline{X^{n-1}}$, $\overline{X^{n-2}}$,
\ldots, $\overline{X^{1}}$, in the form
\begin{equation}
\left. 
\begin{array}{rcl}
\frac{R^{n}}{n!}\frac{d\overline{X^{n}}(t)}{dt} & \simeq  & \frac{R^{n}}{
\left( n-2\right) !}\frac{\overline{X^{n-1}}(t)-\overline{X^{n}}(t)}{4N_{e}}
\\ 
&  &  \\ 
\frac{R^{n-1}}{\left( n-1\right) !}\frac{d\overline{X^{n-1}}(t)}{dt} & = & 
\frac{R^{n}}{\left( n-2\right) !}\frac{\overline{X^{n-1}}(t)-\overline{X^{n}}
(t)}{4N_{e}} \\ 
&  &  \\ 
&  & +\frac{R^{n-1}}{\left( n-3\right) !}\frac{\overline{X^{n-2}}(t)-
\overline{X^{n-1}}(t)}{4N_{e}} \\ 
&  &  \\ 
\frac{R^{n-2}}{\left( n-2\right) !}\frac{d\overline{X^{n-2}}(t)}{dt} & = & 
\frac{R^{n-1}}{\left( n-3\right) !}\frac{\overline{X^{n-2}}(t)-\overline{
X^{n-1}}(t)}{4N_{e}} \\ 
&  &  \\ 
&  & +\frac{R^{n-2}}{\left( n-4\right) !}\frac{\overline{X^{n-3}}(t)-
\overline{X^{n-2}}(t)}{4N_{e}} \\ 
& \vdots  &  \\ 
\frac{R^{1}}{1!}\frac{d\overline{X^{1}}(t)}{dt} & = & R^{2}\frac{\bar{X}(t)-
\overline{X^{2}}(t)}{4N_{e}}.
\end{array}
\right\} 
\end{equation}
It may then be seen that
\begin{equation}
\begin{array}{rcl}
\frac{(-R)^{n}}{n!}\frac{d\overline{X^{n}}(t)}{dt}+\frac{(-R)^{n-1}}{(n-1)!}
\frac{d\overline{X^{n-1}}(t)}{dt}  
...+\frac{(-R)^{1}}{1!}\frac{d\overline{X^{1}}(t)}{dt} & \simeq  & 0
\end{array}
\end{equation} 
and the integral of this equation over $t$, from $0$ to $\infty $ yields $
P_{fix}(y)\simeq \frac{\lbrack 1-e^{-Ry}]_{n}}{[1-e^{-R}]_{n}}$.

%\end{appendixbox}

%\begin{appendixbox}
\section{Solution of the second order approximation with time-dependent
parameters}
In this appendix we present a method for solving the equations for $\bar
{X}(t)$ and $\overline{X^{2}}(t)$ when the quantities $s$ and $N_{e}$ depend
on the time.

We begin with the equations for $\bar{X}(t)$ and $\overline{X^{2}}(t)$ which
now take the form
\begin{align}
\frac{d\bar{X}(t)}{dt}  &  =s(t)\left[  \bar{X}(t)-\overline{X^{2}}(t)\right]
\label{Xbar eq}\\
& \nonumber\\
\frac{d\overline{X^{2}}(t)}{dt}  &  \simeq\frac{1}{2N_{e}(t)}\left[  \bar
{X}(t)-\overline{X^{2}}(t)\right]  . \label{X2bar eq}
\end{align}
and are subject to $\bar{X}(0)=y$ and $\overline{X^{2}}(0)=y^{2}$.

We define
\begin{align}
R(t)  &  =4N_{e}(t)s(t)\\
& \nonumber\\
D(t)  &  =\bar{X}(t)-\overline{X^{2}}(t)\\
& \nonumber\\
\Phi(t)  &  =1-\exp\left(  -\int_{0}^{t}\left(  1-\frac{R(z)}{2}\right)
\frac{dz}{2N_{e}(z)}\right)  .
\end{align}

It follows that $D(t)$ obeys
\begin{eqnarray}
\frac{dD(t)}{dt} &=&-\left( \frac{1}{2N_{e}(t)}-s(t)\right) D(t)  \notag \\
&&  \notag \\
&=&-\frac{1}{2N_{e}(t)}\left( 1-\frac{R(t)}{2}\right) D(t)  \label{D eq}
\end{eqnarray}
and has the solution
\begin{align}
D(t)  &  =D(0)\exp\left(  -\int_{0}^{t}\left(  1-\frac{R(w)}{2}\right)
\frac{dw}{2N_{e}(w)}\right) \nonumber\\
& \nonumber\\
&  =y(1-y)\exp\left(  -\int_{0}^{t}\left(  1-\frac{R(w)}{2}\right)  \frac
{dw}{2N_{e}(w)}\right) \nonumber\\
& \nonumber\\
&  =y(1-y)\left[  1-\Phi(t)\right]  . \label{D explicit}
\end{align}

Proceeding, we rewrite Eq. (\ref{Xbar eq}) as
\begin{equation}
\frac{d\bar{X}(t)}{dt}=\frac{R(t)}{2}\frac{1}{2N_{e}(t)}D(t).
\label{rewrite xbar eq}
\end{equation}
We then use Eqs. (\ref{D eq}) and (\ref{D explicit}) to write
\begin{equation}
\frac{1}{2N_{e}(t)}D(t)=-\frac{1}{1-\frac{R(t)}{2}}\frac{dD(t)}{dt}
=\frac{y(1-y)}{1-\frac{R(t)}{2}}\Phi^{\prime}(t) \label{subs}
\end{equation}
where $\Phi^{\prime}(t)=d\Phi(t)/dt$. Equation (\ref{subs}) allows Eq.
(\ref{rewrite xbar eq}) to be written as
\begin{align}
\frac{d\bar{X}(t)}{dt}  &  =\frac{y(1-y)\frac{R(t)}{2}}{1-\frac{R(t)}{2}}
\Phi^{\prime}(t)=\left(  \frac{\frac{R(t)y}{1!}-\frac{\left[  R(t)y\right]
^{2}}{2!}}{\frac{R(t)}{1!}-\frac{\left[  R(t)\right]  ^{2}}{2!}}-y\right)
\Phi^{\prime}(t)\nonumber\\
& \nonumber\\
&  =\frac{\left[  1-e^{-R(t)y}\right]  _{2}}{\left[  1-e^{-R(t)}\right]  _{2}
}\Phi^{\prime}(t)-y\Phi^{\prime}(t).
\end{align}
On integrating this equation from $0$ to $t$, and using $\bar{X}(0)=y$, we
obtain
\begin{equation}
\bar{X}(t)=\int_{0}^{t}\frac{\left[  1-e^{-R(z)y}\right]  _{2}}{\left[
1-e^{-R(z)}\right]  _{2}}\Phi^{\prime}(z)dz+y\left[  1-\Phi(t)\right]  .
\label{11}
\end{equation}
Using a similar approach, we obtain
\begin{equation}
\overline{X^{2}}(t)=\int_{0}^{t}\frac{\left[  1-e^{-R(z)y}\right]  _{2}
}{\left[  1-e^{-R(z)}\right]  _{2}}\Phi^{\prime}(z)dz+y^{2}\left[
1-\Phi(t)\right]  . \label{22}
\end{equation}

On the assumption that $1-R(t)/2>0$ for all $t$, we have that $\lim
_{t\rightarrow\infty}\Phi(t)=1$ and then both $\bar{X}(t)$ and $\overline
{X^{2}}(t)$ in Eqs. (\ref{11}) and (\ref{22}) have the same long time limit of
$\int_{0}^{\infty}\frac{\left[  1-e^{-R(z)y}\right]  _{2}}{\left[
1-e^{-R(z)}\right]  _{2}}\Phi^{\prime}(z)dz$, which is the second order
approximation of $P_{fix}(y)$.

%\end{appendixbox}

%\begin{appendixbox}
\section{Solution of the third order approximation with time-dependent
parameters}
In this appendix we present a method for solving the equations for $\bar
{X}(t)$, $\overline{X^{2}}(t)$ and $\overline{X^{3}}(t)$, associated with the
third order approximation, when the quantities $s$ and $N_{e}$ depend on the time.

The third order approximation corresponds to solving the equations
\begin{equation}
\left. 
\begin{array}{rcl}
\frac{d\bar{X}(t)}{dt} & = & s\left[ \bar{X}(t)-\overline{X^{2}}(t)\right] 
\\ 
&  &  \\ 
\frac{d\overline{X^{2}}(t)}{dt} & = & \frac{2s}{R}\left[ \bar{X}(t)-
\overline{X^{2}}(t)\right]  \\ 
&  &  \\ 
&  & +2s\left[ \overline{X^{2}}(t))-\overline{X^{3}}(t)\right]  \\ 
&  &  \\ 
\frac{d\overline{X^{3}}(t)}{dt} & \simeq  & \frac{6s}{R}\left[ \overline{
X^{2}}(t)-\overline{X^{3}}(t)\right] 
\end{array}
\right\}   \label{3rd order de}
\end{equation}
(see Appendix 3). However, underlying these three equations are a simpler pair
of coupled equations. In terms of the functions $A(t)$ and $B(t)$ defined by
\begin{equation}
A\equiv A(t)=\bar{X}(t)-\overline{X^{2}}(t),\qquad B\equiv B(t)=\overline
{X^{2}}(t)-\overline{X^{3}}(t)
\end{equation}
we can write
\begin{equation}
\left. 
\begin{array}{rcl}
\dfrac{d\bar{X}}{dt} & = & \dfrac{1}{2N_{e}}\dfrac{R}{2}A \\ 
&  &  \\ 
\dfrac{d\overline{X^{2}}}{dt} & = & \dfrac{1}{2N_{e}}\left( A+RB\right)  \\ 
&  &  \\ 
\dfrac{d\overline{X^{3}}(t)}{dt} & \simeq  & \dfrac{1}{2N_{e}}3B.
\end{array}
\right\}   \label{dx dx2 dx3}
\end{equation}
These equations lead to the pair of coupled equations
\begin{equation}
\left. 
\begin{array}{rcl}
\dfrac{dA}{dt} & = & -\dfrac{1}{2N_{e}}\left[ \left( 1-\dfrac{R}{2}\right)
A+RB\right]  \\ 
&  &  \\ 
\dfrac{dB}{dt} & = & -\dfrac{1}{2N_{e}}\left[ -A+\left( 3-R\right) B\right] 
\end{array}
\right\}   \label{AB system appendix}
\end{equation}
and are subject to $A(0)=y-y^{2}$ and $B(0)=y^{2}-y^{3}$.

We thus need to solve Eq. (\ref{AB system appendix}), for $A(t)$ and $B(t)$,
and statistics of frequencies, can be obtained from knowledge of $A(t)$ and
$B(t)$ by integration. For example from Eq. (\ref{dx dx2 dx3}) we obtain
\begin{equation}
\bar{X}(t)\simeq y+\int_{0}^{t}\frac{R(z)}{4N_{e}(z)}A(z)dz.
\label{Xbar t dep n=3 appendix}
\end{equation}

%\end{appendixbox}

\newpage

%\bibliography{Traj}

\bibliographystyle{ieeetr}
%\bibliography{Traj}
\bibliography{MavreasWaxman_arXiv}

\begin{thebibliography}{10}

\bibitem{Kimura}
M.~Kimura, ``On the probability of fixation of mutant genes in a population,''
  {\em Genetics}, vol.~47, pp.~713--719, 1962.

\bibitem{EwensBook}
W.~Ewens, {\em Mathematical Population Genetics I. Theoretical Introduction,
  2nd Edition}.
\newblock Springer-Verlag, New York, 2004.

\bibitem{Takahata}
N.~Takahata, K.~Ishii, and H.~Matsuda, ``Effect of temporal fluctuation of
  selection coefficient on gene frequency in a population,'' {\em Genetics},
  vol.~72, pp.~4541--4545, 1975.

\bibitem{Takahata_Kimura}
N.~Takahata and M.~Kimura, ``Genetic variability maintained in a finite
  population under mutation and autocorrelated random fluctuation of selection
  intensity,'' {\em Genetics}, vol.~76, pp.~5813--5817, 1979.

\bibitem{Muller}
H.~J. Muller, ``The relation of recombination to mutational advance,'' {\em
  Mutation Research}, vol.~1, pp.~2--9, 1964.

\bibitem{Felsenstein}
J.~Felsenstein, ``The evolutionary advantage of recombination,'' {\em
  Genetics}, vol.~78, p.~737–756, 1974.

\bibitem{Haldane1927}
J.~B.~S. Haldane, ``A mathematical theory of natural and artificial selection,
  part v: Selection and mutation,'' {\em Mathematical Proceedings of the
  Cambridge Philosophical Society}, vol.~23, pp.~838--844, 7 1927.

\bibitem{Mavreas}
K.~Mavreas, T.~I. Gossmann, and D.~Waxman, ``Loss and fixation of strongly
  favoured new variants: Understanding and extending haldane's result via the
  wright-fisher model,'' {\em Biosystems 104759}, 2022.

\bibitem{KimuraOhta_N}
M.~Kimura and T.~Ohta, ``Probability of gene fixation in an expanding finite
  population,'' {\em Proceedings of the National Academy of Sciences}, vol.~71,
  pp.~3377--3379, 1974.

\bibitem{Otto}
S.~P. Otto and M.~C. Whitlock, ``The probability of fixation in populations of
  changing size,'' {\em Genetics}, vol.~146, pp.~723--733, 1997.

\bibitem{Waxman}
D.~Waxman, ``A unified treatment of the probability of fixation when population
  size and the strength of selection change over time,'' {\em Genetics},
  vol.~188, pp.~907--913, 2011.

\bibitem{Lambert}
A.~Lambert, ``Probability of fixation under weak selection: a branching process
  unifying approach,'' {\em Theoretical Population Biology}, vol.~69,
  pp.~419--441, 2006.

\bibitem{Fisher}
R.~A. Fisher, {\em The Genetical Theory of Natural Selection}.
\newblock Clarendon Press, Oxford, 1930.

\bibitem{Wright}
S.~G. Wright, ``Evolution in {M}endelian populations,'' {\em Genetics},
  vol.~16, pp.~97--159, 1931.

\bibitem{Rice}
S.~H. Rice, {\em Numerical Solution of Stochastic Differential Equations}.
\newblock Sinauer Associates: Sunderland, MA, USA, 2004.

\bibitem{ZhaoGossmannWaxman}
L.~Zhao, T.~I. Gossmann, and D.~Waxman, ``A modified wright–fisher model that
  incorporates ne: A variant of the standard model with increased biological
  realism and reduced computational complexity,'' {\em Journal of Theoretical
  Biology}, vol.~393, pp.~218–--228, 2016.

\bibitem{McKane_Waxman}
A.~McKane and D.~Waxman, ``Singular solutions of the diffusion equation of
  population genetics,'' {\em Journal of Theoretical Biology}, vol.~247,
  pp.~849--858, 2007.

\bibitem{Tuckwell}
H.~Tuckwell, {\em Elementary Applications of Probability Theory: With an
  introduction to stochastic differential equations}.
\newblock Chapman and Hall, London, 1979.

\bibitem{Ito}
K.~Ito, ``Stochastic integral,'' {\em Proc. Imperial Acad.}, vol.~20,
  pp.~519--524, 1944.

\bibitem{Kloeden}
P.~E. Kloeden and E.~Platen, {\em Numerical Solution of Stochastic Differential
  Equations}.
\newblock Springer, 1992.

\end{thebibliography}

\end{document}